\documentclass[conference]{IEEEtran}  
\usepackage{graphicx}
\usepackage{amsmath}
\usepackage{amssymb}
\usepackage{booktabs}
\usepackage{enumerate}
\usepackage{algorithm}
\usepackage{algorithmic}
\usepackage{amsmath}
\usepackage{graphics}
\usepackage{epsfig}
\usepackage{subfigure}
\usepackage{mmap} 

\usepackage{bbding}
\usepackage{amsmath}
\usepackage{amssymb}
\usepackage{ntheorem}
\usepackage{graphicx}
\usepackage{xcolor}

\psfull

\begin{document}
\title{AN-aided Secure Transmission in Multi-user MIMO SWIPT Systems}
\author{\IEEEauthorblockN{Zhengyu Zhu$^{\dag}$,  Ning Wang$^\dag$, 
Zheng Chu$^\S$, Zhongyong Wang$^\dag$,  Inkyu Lee$^ \ddag$\\}
\IEEEauthorblockN{$^\dag$ School of Information Engineering, Zhengzhou University, China \\}
\IEEEauthorblockN{$^\S$ 5G Innovation Center (5GIC), Institute of Communication
Systems (ICS), University of Surrey, UK\\}
\IEEEauthorblockN{$^\ddag$ School of Electrical Engineering, Korea University, Korea}
Email: zhuzhengyu6@gmail.com, ienwang@zzu.edu.cn, iezywang@zzu.edu.cn,  inkyu@korea.ac.kr}

\setlength{\abovedisplayskip}{4pt}
\setlength{\belowdisplayskip}{4pt}

\maketitle
\begin{abstract}
In this paper, an energy harvesting scheme for a multi-user multiple-input-multiple-output (MIMO) secrecy channel with artificial noise (AN) transmission is investigated. Joint optimization of the transmit beamforming matrix, the AN covariance matrix, and the power splitting ratio is conducted to minimize the transmit power under the target secrecy rate, the total transmit power, and the harvested energy constraints.
The original problem is shown to be non-convex, which is tackled by a two-layer decomposition approach.
The inner layer problem is solved through semi-definite relaxation, and the outer problem is shown to be a single-variable optimization that can be solved by one-dimensional (1-D) line search.
To reduce computational complexity, a sequential parametric convex approximation (SPCA) method is proposed to find a near-optimal solution.
Furthermore, tightness of the relaxation for the 1-D search method is validated by showing that the optimal solution of the relaxed problem is rank-one.
Simulation results demonstrate that the proposed SPCA method achieves the same performance as the scheme based on 1-D search method but with much lower complexity.
\end{abstract}

\IEEEpeerreviewmaketitle

%

\section{Introduction}
In recent years, the idea of energy harvesting (EH) has been introduced to power electronic devices by energy captured from
the environment. \cite{1EH}.
Based on this idea, simultaneous wireless information and power transfer (SWIPT) schemes have been proposed to extend the lifetime
of wireless networks \cite{5ZhangR_13TWC_MIMO_SWIPT}-\cite{9zhu16_JCN}.
For SWIPT operation in multiple antenna systems \cite{8zhu_16VTC}, co-located receiver architecture employing a power splitter for EH and information decoding (ID) has been studied \cite{9zhu16_JCN}.

On the other hand, in the literature we see increasing research interest in secrecy transmission through physical layer (PHY) security designs \cite{11chutwc} \cite{zhu_secure_SWIPT}.
By adding artificial noise (AN) and projecting it onto the null space of information user channels, the eavesdroppers would experience a higher noise floor and thus obtain less information about the messages transmitted to the legitimate receivers \cite{14Negi_08TWC_tichu_AN}.
The AN-aided beamforming for SWIPT operation has been investigated in various multiple-input multiple-output (MIMO) channels \cite{16LiuL_14TSP_Secrecy_WIPT_MISO}-\cite{21Chu_SWIPT_MISO_secrecy}.

In SWIPT systems, a new information security issue is raised because the energy-harvesting receivers (ERs) can potentially eavesdrop the information transmission to the  information receivers (IRs) with relatively higher received signal strength \cite{16LiuL_14TSP_Secrecy_WIPT_MISO}.
In order to guarantee information security for the IRs, it is desirable to implement some mechanism to prevent the ERs from recovering the confidential message from their observations.


In this paper, we study a AN-aided secrecy transmission over a multi-user MIMO secrecy channel which consists of one multi-antenna transmitter, multiple legitimate single antenna co-located receivers (CRs) with power splitter and multiple multi-antenna ERs.
The design objective is to jointly optimize the transmit beamforming matrix, the AN covariance matrix, and the power splitting (PS) ratio
such that the AN transmit power is maximized\footnote{AN power maximization is equivalent to minimizing the transmit power of the information signal \cite{18Shi_Secrecy_MISO_BC_SWIPT}.} subject to constraints on the secrecy rate, the total transmit power, and the energy harvested by both the CRs and the ERs.
Because of the coupling effect in the joint optimization problem, determination of the AN covariance matrix and the PS ratio makes the derivation of the secrecy rate and the harvested energy at the CRs more complicated. 

The formulated power minimization (PM) problem for AN-aided secrecy transmission is shown to be non-convex, which cannot be solved directly \cite{26Boyd_Convex}.
First, the PM problem is thus transformed into a two-layer optimization problem.
The inner loop of the PM problem is solved through semi-definite relaxation (SDR), 
while the outer loop is shown to be a single-variable optimization problem, where a one-dimensional (1-D) line search algorithm 
is employed to find the optima. To reduce computational complexity, a sequential parametric convex approximation (SPCA) method is also investigated \cite{SPCA_method}. The tightness of the SDR is verified by showing that the optimal solution is rank-one.


\textbf{\emph{Notation:}}  ${{\emph{vec}}}(\textbf{{A}})$ stacks the elements of $\textbf{{A}}$
    in a column vector. $\mathbf{0}_{M\times L}$ is a zero matrix of size ${M\times L}$. $\mathrm{E}\{\cdot\}$ is the expectation operator, and $\Re\{\cdot\}$ stands for the real part of a complex number.
     $[x]^{+}$ represents $\max\{x,0\}$ and $\lambda_{max}(\textbf{{A}})$ denotes the maximum eigenvalue of $\textbf{{A}}$.


\section{System Model}
In this section, we consider a multi-user MIMO secrecy channel which consists of one multi-antenna transmitter,
$ L $ single-antenna CRs and $ K $ multi-antenna ERs. We assume that each CR employs the PS scheme to receive the information and harvest power simultaneously.
It is assumed that the transmitter is equipped with $ N_{T} $ transmit antennas, and each ER has $ N_{R} $ receive antennas.

We denote by $ \mathbf{h}_{c,l} \in \mathbb{C}^{N_{T}} $ the channel vector between the transmitter and the $ l $-th CR, and $ \mathbf{H}_{e,k} \in  \mathbb{C}^{N_{T} \times N_{R}} $ the channel matrix between the transmitter and the $ k $-th ER.
The received signal at the $ l $-th CR and the $ k $-th ER are given by
\begin{equation*}
\begin{split}
y_{c,l} & = \mathbf{h}_{c,l}^{H}\mathbf{x} \!+\! n_{c,l}, ~\forall l,\\
\mathbf{y}_{e,k} & = \mathbf{H}_{e,k}^{H}\mathbf{x}\!+\! \mathbf{n}_{e,k}, ~\forall k,
\end{split}
\end{equation*}
where $ \mathbf{x} \in \mathbb{C}^{N_{T}} $ is the transmitted signal vector, and $ n_{c,l} \sim \mathcal{CN}(0, \sigma_{c,l}^{2}) $
and $ \mathbf{n}_{e,k} \sim \mathcal{CN}(0, \sigma_{k}^{2}\mathbf{I}) $ are the additive Gaussian noise at the $ l $-th CR and the $ k $-th ER, respectively.

In order to achieve secure transmission, the transmitter employs
transmit beamforming with AN, which acts as interference to the ERs, and provides energy to the CRs and ERs.
The transmit signal vector $ \mathbf{x} $ can be written as
\begin{equation}
\mathbf{x} = \mathbf{q}s + \mathbf{w},
\end{equation}
where $ \mathbf{q} \in \mathbb{C}^{N_{T}}$ defines the transmit beamforming vector, $s$ with $\mathrm{E}\{s^2\} = 1$ is the information-bearing signal intended for the CRs, and $ \mathbf{w} \in \mathbb{C}^{N_{T}}$ represents the energy-carrying AN, which can also be composed by multiple energy beams.

As the CR adopts PS to perform ID and EH simultaneously,
the received signal at the $l$-th CR is divided into ID and EH components by the PS ratio $\rho_{c,l}\in (0, 1]$.
Therefore, the signal for information detection at the $l$-th CR is given by
\begin{equation*}
y_{c,l}^{ID}   = \sqrt{\rho_{c,l}}y_{c,l} \!+\! n_{p,l}
 =  \sqrt{\rho_{c,l}}(\mathbf{h}_{c,l}^{H}\mathbf{x} \!+\! n_{c,l}) \!+\! n_{p,l}, ~\forall l,
\end{equation*}
where $ n_{p,l} \sim \mathcal{CN}(0, \sigma_{p,l}^{2}) $  is the additive Gaussian noise at the  $l$-th CR.

Denoting $ \mathbf{Q} = \mathrm{E}\{\mathbf{q}\mathbf{q}^H\} $ as the transmit covariance matrix and
$ \mathbf{W}= \mathrm{E}\{\mathbf{w}\mathbf{w}^H\} $ as the AN covariance matrix,
the achieved secrecy rate at the $ l $-th CR is given by
\begin{equation}\label{eq:Achieved_sec_rate_with_user_PS}
\begin{split}
\hat{R}_{c,l} &= \bigg[\log\bigg( 1 \!+\! \frac{\rho_{c,l} \mathbf{h}_{c,l}^{H}\mathbf{Q}\mathbf{h}_{c,l}}{\rho_{c,l}
(\sigma_{c,l}^{2}\!+\!\mathbf{h}_{c,l}^{H}\mathbf{W}\mathbf{h}_{c,l}) \!+\! \sigma_{p,l}^{2}} \bigg)\\
&- \max_{k} \log \bigg| \mathbf{I} \!+\! (\mathbf{H}_{e,k}^{H}\mathbf{W}\mathbf{H}_{e,k} \!+\!
 \sigma_{k}^{2}\mathbf{I})^{-1} \mathbf{H}_{e,k}^{H}\mathbf{Q}\mathbf{H}_{e,k} \bigg| \bigg]^{+},~\forall l.
\end{split}
\end{equation}
The harvested power at the $ l $-th CR and the $ k $-th ER is therefore
\begin{equation}
\begin{split}
E_{c,l} & = \eta_{c,l}(1 \!-\! \rho_{c,l})\big(\mathbf{h}_{c,l}^{H}(\mathbf{Q}\!+\!\mathbf{W})\mathbf{h}_{c,l} \!+\! \sigma_{c,l}^{2}\big), ~\forall l,\\
E_{e,k} & = \eta_{e,k}\bigg(\textrm{tr}\big(\mathbf{H}_{e,k}^{H}(\mathbf{Q}\!+\!\mathbf{W})\mathbf{H}_{e,k}\big) \!+\! N_{R}\sigma_{k}^{2}\bigg), ~\forall k,
\end{split}
\end{equation}
where $ \eta_{c,l} $ and $ \eta_{e,k} $ represent the EH efficiency of the $ l $-th CR and
the EH efficiency of the $ k $-th ER, respectively. In this paper we set $ \eta_{c,l}=\eta_{e,k}=0.3 $ for simplicity.


\section{Masked Beamforming Based Power Minimization}
In this section, we study transmit beamforming optimization under the assumption that perfect CSI of all the channels is available at the transmitter.
\subsection{Problem Formulation}
In this problem, the transmit power of the information signal is minimized subject to the
 total transmit power constraint, the secrecy rate constraint, and the EH constraints of the CRs and the ERs
 such that the AN transmit power is maximized for secrecy consideration.
The AN-aided PM problem is thus formulated as
\begin{subequations}\label{eq:Masked_beamforming_sec_rate_opt_ori}
\begin{eqnarray}
\min_{\mathbf{Q},{\kern 1pt}\mathbf{W},{\kern 1pt}\rho_{c,l}} &&~~~~~~~~~ \textrm{tr}(\mathbf{Q}) \nonumber\\
&&\!\!\!\!\!\!\!\!\!\!\!\!\!\!\!\!\!\!\!\!\!\!\!\!\!\!\!\!\!\!\!\!\!\!\! \mbox{s.t.} ~~ \hat{R}_{c,l} \!\geq\! \bar{R}_{c,l}, \forall l,
\label{eq:Achieved_sec_rate_constraint}\\
&&\!\!\!\!\!\!\!\!\!\!\!\!\!\!\!\!\!\!\!\!\!\!\! \textrm{tr}(\mathbf{Q} + \mathbf{W}) \leq P,  \label{eq:Power_constraints}\\
&&\!\!\!\!\!\!\!\!\!\!\!\!\!\!\!\!\!\!\!\!\!\!\! \mathbf{h}_{c,l}^{H}(\mathbf{Q} + \mathbf{W})\mathbf{h}_{c,l} + \sigma_{c,l}^{2} \geq \frac{\bar{E}_{c,l}}{\eta_{c,l}(1-\rho_{c,l})}, \label{eq:Energy_constraint_user_PS}\\
&&\!\!\!\!\!\!\!\!\!\!\!\!\!\!\!\!\!\!\!\!\!\!\!  \min_{k} \textrm{tr}\big(\mathbf{H}_{e,k}^{H}(\mathbf{Q}\!+\!\mathbf{W})\mathbf{H}_{e,k}\big) \!+\! N_{R}\sigma_{k}^{2}  \geq \frac{\bar{E}_{e,k}}{\eta_{e,k}},\forall k, \label{eq:Energy_constraint_eve}\\
&&\!\!\!\!\!\!\!\!\!\!\!\!\!\!\!\!\!\!\!\!\!\!\! \mathbf{Q}\succeq \mathbf{0},\mathbf{W} \succeq \mathbf{0}, 0 < \rho_{c,l} \leq 1, \forall l,  \textrm{rank}(\mathbf{Q}) \!=\! 1, \label{eq:Another_constraints_ori}
\end{eqnarray}
\end{subequations}
where $ \bar{R}_{c,l} $ is the target secrecy rate, $ P $ is the total transmit power,
and $ \bar{E}_{c,l} $ and $ \bar{E}_{e,k} $ denote the predefined harvested power at the $l$-th CR and
the $k$-th ER, respectively.
The constraint (\ref{eq:Energy_constraint_eve}) guarantees that a minimum energy harvested power should be achieved by the $k$-th ER.
\subsection{One-Dimensional Line Search Method (1-D Search)}
Problem \eqref{eq:Masked_beamforming_sec_rate_opt_ori} is non-convex due to the
 secrecy rate constraint (\ref{eq:Achieved_sec_rate_constraint}), and thus cannot be solved directly.
 In order to circumvent this issue, we convert the original problem
by introducing a slack variable $ t $ for the $ k $-th ER's rate. Then we have
\begin{subequations}\label{eq:relaxed_power_min_problem1}
\begin{eqnarray}
&&\!\!\!\!\!\!\!\!\!\!\!\!\!\!\! \min_{\mathbf{Q},{\kern 1pt}\mathbf{W},{\kern 1pt}\rho_{c,l},{\kern 1pt}t}  ~~~~~~ \textrm{tr}(\mathbf{Q}) \nonumber \\
&&\!\!\!\!\!\!\!\!\!\!\!\!\!\!\! \mbox{s.t.}~~ \!  \log\bigg(t \!+\! \frac{t\rho_{c,l}\mathbf{h}_{c,l}^{H}\mathbf{Q}\mathbf{h}_{c,l}}{\rho_{c,l}(\sigma_{c,l}^{2}
\!+\!\mathbf{h}_{c,l}^{H}\mathbf{W}\mathbf{h}_{c,l})\!+\!\sigma_{p,l}^{2}}\bigg) \!\geq\! \bar{R}_{c,l}, \forall l,\label{eq:objective_function_with_log}
\\
&&\!\!\!\!\!\!\!  \bigg| \mathbf{I} \!+\! (\sigma_{k}^{2}\mathbf{I} \!+\! \mathbf{H}_{e,k}^{H}\mathbf{W}\mathbf{H}_{e,k})^{-1}\mathbf{H}_{e,k}^{H}\mathbf{Q}\mathbf{H}_{e,k} \bigg|
 \! \leq \! \frac{1}{t},\forall k,  \label{eq:Eavesdroppers_rate_slack_variable_log}  \\
&&\!\!\!\!\!\!\! \eqref{eq:Power_constraints}-\eqref{eq:Another_constraints_ori}. \nonumber
\end{eqnarray}
\end{subequations}

Problem \eqref{eq:relaxed_power_min_problem1} is still non-convex in constraints
\eqref{eq:objective_function_with_log} and \eqref{eq:Eavesdroppers_rate_slack_variable_log},
which can be addressed by reformulating \eqref{eq:relaxed_power_min_problem1} into a two-layer problem.
For the inner layer, we solve problem \eqref{eq:relaxed_power_min_problem1} for a given $ t $, which
is relaxed as
\begin{subequations}\label{eq:relaxed_power_min_problem_results}
\begin{eqnarray}
&&\!\!\!\!\!\!\!\!\!\!\!\!\!\!\!\!\!\!\!\!\!\! f(t) = \min_{\mathbf{Q},{\kern 1pt}\mathbf{W},{\kern 1pt}\rho_{c,l},{\kern 1pt}t}  ~~  \textrm{tr}(\mathbf{Q}) \nonumber\\
&&\!\!\!\!\!\!\!\!\!\!\!\!\!\!\!\!\!\!\!\!\!\!  \mbox{s.t.}~    \mathbf{h}_{c,l}^{H}\big( t \mathbf{Q} \!-\! (2^{\bar{R}_{c,l}} \!-\! t)\mathbf{W} \big) \mathbf{h}_{c,l} \!\geq\!
(2^{\bar{R}_{c,l}} \!-\! t)\big(\sigma_{c,l}^{2}\!+\!\frac{\sigma_{p,l}^{2}}{\rho_{c,l}}\big), \label{eq:objective_function_with_log_relaxed}\\
&&\!\!\!\!   ( {\textstyle{1 \over t}} \!-\! 1)(\sigma_{k}^{2}\mathbf{I} \!+\! \mathbf{H}_{e,k}^{H}\mathbf{W}\mathbf{H}_{e,k}) \!\succeq\!
\mathbf{H}_{e,k}^{H}\mathbf{Q}\mathbf{H}_{e,k},\forall k, \label{eq:Eavesdroppers_rate_slack_variable_log_relaxed}\\
&&\!\!\!\!   \eqref{eq:Power_constraints}-\eqref{eq:Another_constraints_ori}, \nonumber
\end{eqnarray}
\end{subequations}
where $ f(t) $ is defined as  the optimal value of problem \eqref{eq:relaxed_power_min_problem_results}, which is a function of $ t $.
Even though the function $f(t)$ cannot be expressed in closed-form, numerical evaluation of $f(t)$ is feasible. Note that the LMI constraint \eqref{eq:Eavesdroppers_rate_slack_variable_log_relaxed} is obtained from \cite[Proposition 1]{15LiQ_13TSP_Spatially_selective_AN}, and is based on the assumption that $ \textrm{rank}(\mathbf{Q}) \leq 1 $, which will be shown later.

%
%
%
%
%

By ignoring the non-convex constraint $ \textrm{rank}(\mathbf{Q}) = 1 $, problem \eqref{eq:relaxed_power_min_problem_results} becomes convex and thus can be solved efficiently by an interior-point method for any given $ t $ \cite{26Boyd_Convex}.
The outer layer problem, whose objective is to find the optimal value of $ t $, is then formulated as
\begin{equation}\label{eq:Outer_level_problem_over_t}
\begin{split}
 \min_{t} ~ f(t) ~~~~~~~~ \mbox{s.t.} ~~ t_{\textrm{min}} \!\leq\! t \!\leq\! t_{\textrm{max}},
\end{split}
\end{equation}
where $t_{\textrm{max}}$ and $t_{\textrm{min}}$ are the upper and lower bounds of $t$, respectively.
The solution to problem \eqref{eq:Outer_level_problem_over_t} can be found by one-dimensional line search. 
For the line search algorithm, we need to determine the lower and upper bounds of the searching interval for $ t $.
It is straightforward that $ t_{\textrm{max}} = 1 $ can be used as the upper bound due to the feasibility of \eqref{eq:Eavesdroppers_rate_slack_variable_log}, while a lower bound is calculated as
\begin{eqnarray}
t &&\geq \min_{l}  \bigg(1 + \frac{\rho_{c,l} \mathbf{h}_{c,l}^{H}\mathbf{Q}\mathbf{h}_{c,l}}{\rho_{c,l} (\sigma_{c,l}^{2}+\mathbf{h}_{c,l}^{H}\mathbf{W}\mathbf{h}_{c,l}) + \sigma_{p,l}^{2}} \bigg)^{-1} \nonumber\\
 &&\geq  \min_{l} \bigg(1 +\frac{\mathbf{h}_{c,l}^{H}\mathbf{Q}\mathbf{h}_{c,l}}{\sigma_{c,l}^{2}+\sigma_{p,l}^{2}+\mathbf{h}_{c,l}^{H}\mathbf{W}\mathbf{h}_{c,l}} \bigg)^{-1} \\
 &&\geq  \min_{l} \bigg(1+\frac{P \|\mathbf{h}_{c,l}\|^{2}}{\sigma_{c,l}^{2}+\sigma_{p,l}^{2}}\bigg)^{-1} \triangleq t_{\textrm{min}}\nonumber.
\end{eqnarray}
where the first inequality is based on the secrecy rate $\bar{R}_{c,l} \geq 0$, the third inequality
follows from \eqref{eq:Power_constraints}.

Utilizing the results in Remark 1, next we show the tightness of the AN-aided PM problem \eqref{eq:Masked_beamforming_sec_rate_opt_ori}
by the following theorem.

\emph{\underline{Theorem 1:}} Provided that problem
\eqref{eq:relaxed_power_min_problem_results} is feasible for a given $ t > 0 $, there
exists an optimal solution to \eqref{eq:Masked_beamforming_sec_rate_opt_ori} such that the rank of $ \mathbf{Q} $ is always equal to 1.

\emph{\underline{Proof:} } See Appendix A. ~~~~~~~~~~~~~~~~~~~~~~~~~~~~~~~~~~~~~~~~~~~~~~~~~~~~~~~$\blacksquare $

Problem \eqref{eq:Outer_level_problem_over_t} can be solved by conducting
one-dimensional line search for $f(t)$ over $t$ and choosing the minimum $f(t)$ as the optimal solution. 
Solving the SDP problem \eqref{eq:relaxed_power_min_problem_results} with the optimal $f(t)$, we can obtain the optimal design variables $(\mathbf{Q}^*,{\kern 1pt}\mathbf{W}^*,{\kern 1pt}\rho_{c,l}^*)$. The optimal beamforming vector $\mathbf{q}^*$ is then computed by eigenvalue decomposition $\mathbf{Q}^* = \mathbf{q}^*\mathbf{q}^{*H}$.
\subsection{Low-Complexity SPCA Algorithm}
In this subsection, we propose an SPCA based iterative method to reduce the computational complexity.
By introducing two slack variables $r_1 > 0$ and $r_2 > 0$, the constraint \eqref{eq:Achieved_sec_rate_constraint} can be rewritten as
\begin{subequations}\label{eq:9}
\begin{eqnarray}
&&\!\!\!\!\!\!\!\!\!\!\!\!\!\!  \log( r_1r_2 ) \geq  \bar{R}_{c,l}, \forall l,\label{eq:9a}\\
&&\!\!\!\!\!\!\!\!\!\!\!\!\!\!  1 \!+\! \frac{\rho_{c,l} \mathbf{h}_{c,l}^{H}\mathbf{Q}\mathbf{h}_{c,l}}{\rho_{c,l} (\sigma_{c,l}^{2}\!+\!\mathbf{h}_{c,l}^{H}\mathbf{W}\mathbf{h}_{c,l}) \!+\!
\sigma_{p,l}^{2}}  \geq r_1, \forall l, \label{eq:9b}\\
&&\!\!\!\!\!\!\!\!\!\!\!\!\!\! 1 + \frac{{\rm{tr}}({{{{{{ {\mathbf{H}}}}^H_{e,k}}}}
 {{{\mathbf{Q}}}}{{{ {\mathbf{H}}}}_{e,k}}})}
 {{\sigma _k^2 + {\rm{tr}}({{{{{ {\mathbf{H}}}}^H_{e,k}}}}{\mathbf{W}}{{{ {\mathbf{H}}}}_{e,k}}})}  \leq \frac{1}{r_2}, \forall k, \label{eq:9c}
\end{eqnarray}
\end{subequations}
which can be further simplified as
\begin{subequations} \label{eq:10}
\begin{eqnarray}
&&\!\!\!\!\!\!\!\!\!\!\!\!\!\!   r_1r_2  \geq  2^{\bar{R}_{c,l}}, \forall l, \label{eq:10a}\\
&&\!\!\!\!\!\!\!\!\!\!\!\!\!\!   \frac{ \mathbf{h}_{c,l}^{H}\mathbf{Q}\mathbf{h}_{c,l}}{ \sigma_{c,l}^{2}\!+\!\mathbf{h}_{c,l}^{H}\mathbf{W}\mathbf{h}_{c,l} \!+\!
\frac{\sigma_{p,l}^{2}}{\rho_{c,l}}}  \geq {r_1} -1, \forall l,\label{eq:10b}\\
&&\!\!\!\!\!\!\!\!\!\!\!\!\!\!  \frac{{\sigma _k^2 + {{\rm{tr}}({{{{{{ {\mathbf{H}}}}^H_{e,k}}}}
 {{{\mathbf{W}}}}{{{ {\mathbf{H}}}}_{e,k}}})}}}
 {{{\sigma _k^2}}+{{\rm{tr}}({{{{{{ {\mathbf{H}}}}^H_{e,k}}}}
 ({{{\mathbf{Q}}}}+{{{\mathbf{W}}}}){{{ {\mathbf{H}}}}_{e,k}}})}}  \geq  {{r_2}}, \forall k. \label{eq:10c}
\end{eqnarray}
\end{subequations}
The inequality constraint \eqref{eq:10a} is equivalent to ${2^{\bar{R}_{c,l}+2}} + (r_1-r_2)^2 \leq (r_1+r_2)^2$,
which can be converted into a conic quadratic-representable function form as
\begin{equation} \label{eq:11a}
\left\|\left[\sqrt{2^{\bar{R}_{c,l}+2}}~~~~ r_1-r_2 \right]\right\|\leq r_1+r_2, \forall l.
\end{equation}

By transforming inequality constraints \eqref{eq:10b} and \eqref{eq:10c} into
\begin{subequations}\label{eq:13}
\begin{eqnarray}
&&\!\!\!\!\!\!\!\!   {\sigma_{c,l}^{2}\!+\!\mathbf{w}^{H}\mathbf{H}_{c,l}\mathbf{w} \!+\!
\frac{\sigma_{p,l}^{2}}{\rho_{c,l}}} \leq \frac{\mathbf{q}^H{\mathbf{H}_{c,l}}\mathbf{q}}{r_1 -1}, ~\forall l, \label{eq:13a}\\
&&\!\!\!\!\!\!\!\!\!\!\!\!\!\!\!\!\!\!\!\!\!\!\!\!\!   {{{\sigma _k^2}}+ \mathbf{w}^{H}\mathbf{\hat{H}}_{e,k}\mathbf{w} +\mathbf{q}^{H}\mathbf{\hat{H}}_{e,k}\mathbf{q} } \leq \frac{{{\sigma _k^2 + \mathbf{w}^{H}\mathbf{\hat{H}}_{e,k}\mathbf{w}}}}{r_2}, ~\forall k,\label{eq:13b}
\end{eqnarray}
\end{subequations}
where $\mathbf{H}_{c,l} = {\mathbf{h}_{c,l}\mathbf{h}_{c,l}^{H}}$ and $\mathbf{\hat{H}}_{e,k} = {\mathbf{H}_{e,k}\mathbf{H}_{e,k}^{H}}$,
we observe that these two constraints are non-convex, but the right-hand side (RHS) of both (\ref{eq:13a}) and (\ref{eq:13b}) have the function form of quadratic-over-linear, which are convex functions \cite{26Boyd_Convex}.
Based on the idea of the constrained convex procedure \cite{SPCA_firstorder},
these quadratic-over-linear functions can be replaced by their first-order expansions, which transforms the problem into convex programming.
Specifically, we define
\begin{eqnarray}\label{eq:zzy1}
f_{\mathbf{{A}},a}(\mathbf{w},t) = \frac{{{\mathbf{w}^{H}\mathbf{{A}}\mathbf{w}}}}{t-a},
\end{eqnarray}
where $\mathbf{{A}} \succeq \textbf{0}$ and $t \geq a$. At a certain point $(\mathbf{\tilde{w}}, \tilde{t})$, the first-order Taylor expansion of \eqref{eq:zzy1}
 is given by
\begin{eqnarray}\label{eq:zzy2}
F_{\mathbf{{A}},a}(\mathbf{w},t,\mathbf{\tilde{w}},\tilde{t}) = \frac{2\Re{\{\mathbf{\tilde{w}}^{H}\mathbf{{A}}\mathbf{w}\}}}{\tilde{t}-a} - \frac{{{\mathbf{\tilde{w}}^{H}\mathbf{{A}}\mathbf{\tilde{w}}}}}{(\tilde{t}-a)^2}(t-a).
\end{eqnarray}

By using the above results of Taylor expansion, for the points $(\mathbf{\tilde{q}}, \tilde{r}_1)$ and $(\mathbf{\tilde{w}}, \tilde{r}_2)$, we can transform constraints \eqref{eq:13a} and \eqref{eq:13b} into convex forms,  respectively, as
\begin{subequations}\label{eq:zzy3}
\begin{eqnarray}
&&\!\!\!\!\!\!\!\!\!\!\!\!\!\!\!\!\!\!\!    {\sigma_{c,l}^{2}+\mathbf{w}^{H}\mathbf{H}_{c,l}\mathbf{w} +
\frac{\sigma_{p,l}^{2}}{\rho_{c,l}}} \leq F_{{\mathbf{H}_{c,l}},1}(\mathbf{q},r_1,\mathbf{\tilde{q}},\tilde{r}_1), \forall l, \label{eq:zzy3a}\\
&&\!\!\!\!\!\!\!\!\!\!\!\!\!\!\!\!\!\!\!   {{{\sigma _k^2}}+ \mathbf{w}^{H}\mathbf{\hat{H}}_{e,k}\mathbf{w} + \mathbf{q}^{H}\mathbf{\hat{H}}_{e,k}\mathbf{q} } \leq  \sigma _k^2(\frac{2}{\tilde{r}_2} - \frac{r_2}{\tilde{r}^2_2}) \nonumber \\
&&~~~~~~~~~~~~~~~~~~~  +  F_{{\mathbf{\hat{H}}_{e,k}},0}(\mathbf{w},r_2,\mathbf{\tilde{w}},\tilde{r}_2), \forall k.\label{eq:zzy3b}
\end{eqnarray}
\end{subequations}
Denoting $g_{r_1,l} = F_{{\mathbf{H}_{c,l}},1}(\mathbf{q},r_1,\mathbf{\tilde{q}},\tilde{r}_1) -\sigma_{c,l}^{2}- \frac{\sigma_{p,l}^{2}}{\rho_{c,l}}$ and $g_{r_2,k} =  \sigma _k^2(\frac{2}{\tilde{r}_2}\! -\! \frac{r_2}{\tilde{r}^2_2})\! + \! F_{{\mathbf{\hat{H}}_{e,k}},0}(\mathbf{w},r_2,\mathbf{\tilde{w}},\tilde{r}_2) - {{\sigma _k^2}}$, 
\eqref{eq:zzy3a} and \eqref{eq:zzy3b} can be recast as the following second-order cone (SOC) constraints
\begin{subequations}\label{eq:zzy8}
\begin{eqnarray}
&&\!\!\!\!\!\!\!\!\!\!\!\!\!\!\!\!   \big\| [2\mathbf{w}^{H}\mathbf{h}_{c,l}, g_{r_1,l} -1  ]^T \big\|\leq g_{r_1,l} +1, ~\forall l, \label{eq:zzy8a}\\
&&\!\!\!\!\!\!\!\!\!\!\!\!\!\!\!\!\!\!\!\!\!\!\!\!\!    \big\| [2\mathbf{w}^{H}\mathbf{{H}}_{e,k}; 2\mathbf{q}^{H}\mathbf{{H}}_{e,k}; g_{r_2,k} -1]^T \big\| \leq  g_{r_2,k} +1, ~\forall k.\label{eq:zzy8b}
\end{eqnarray}
\end{subequations}
Next we employ the SPCA technique for the SOC constraints \eqref{eq:Energy_constraint_user_PS} and \eqref{eq:Energy_constraint_eve} \cite{Complexity_zhu} to obtain convex approximations.
By substituting $\mathbf{q} \triangleq {\mathbf{\tilde{q}}}+\Delta \mathbf{q}$ and $\mathbf{w} \triangleq {\mathbf{\tilde{w}}}+\Delta \mathbf{w}$  into the left-hand side (LHS) of \eqref{eq:Energy_constraint_user_PS} and \eqref{eq:Energy_constraint_eve}, we obtain
\begin{equation}\label{eq: zzy4}
\begin{split}
&~~~~\mathbf{q}^H\mathbf{H}_{c,l}\mathbf{q}+\mathbf{w}^H\mathbf{H}_{c,l}\mathbf{w} + {{\sigma _k^2}} \\
&=  ({\mathbf{\tilde{q}}}\!+\!\Delta \mathbf{q})^H\mathbf{H}_{c,l}({\mathbf{\tilde{q}}}\!+\!\Delta \mathbf{q})\!+\!({\mathbf{\tilde{w}}}\!+\!\Delta \mathbf{w})^H\mathbf{H}_{c,l}({\mathbf{\tilde{w}}}\!+\!\Delta \mathbf{w})+{{\sigma _k^2}} \\
& \ge {\mathbf{\tilde{q}}}^H\mathbf{H}_{c,l}{\mathbf{\tilde{q}}} + 2\Re \{ {\mathbf{\tilde{q}}}^H\mathbf{H}_{c,l}\Delta {\mathbf{q}}\}  + {\mathbf{\tilde{w}}}^H\mathbf{H}_{c,l}{\mathbf{\tilde{w}}} \\
&~~~~ + 2\Re \{ {\mathbf{\tilde{w}}}^H\mathbf{H}_{c,l}\Delta {\mathbf{w}}\}+{{\sigma _k^2}},
\end{split}
\end{equation}
where the inequality is given by dropping the quadratic terms $\Delta \mathbf{q}^H \mathbf{H}_{c,l}\Delta \mathbf{q}$
 and $\Delta \mathbf{w}^H \mathbf{H}_{c,l}\Delta \mathbf{w}$.
Similarly, in the LHS of \eqref{eq:Energy_constraint_eve}, we have
\begin{equation}\label{eq: zzy5}
\begin{split}
\!\!\!\!\! &\textrm{tr}\big(\mathbf{H}_{e,k}^{H}(\mathbf{Q}\!+\!\mathbf{W})\mathbf{H}_{e,k}\big)\\
\!\!\!\!\! =& \mathbf{q}^H\mathbf{\hat{H}}_{e,k}\mathbf{q}+\mathbf{w}^H\mathbf{\hat{H}}_{e,k}\mathbf{w} \\
\!\!\!\!\! =&  ({\mathbf{\tilde{q}}}\!+\!\Delta \mathbf{q})^H\mathbf{\hat{H}}_{e,k}({\mathbf{\tilde{q}}}\!+\!\Delta \mathbf{q})\!+\!({\mathbf{\tilde{w}}}\!+\!\Delta \mathbf{w})^H\mathbf{\hat{H}}_{e,k}({\mathbf{\tilde{w}}}\!+\!\Delta \mathbf{w}) \\
\!\!\!\!\! \ge &  {\mathbf{\tilde{q}}}^H\mathbf{\hat{H}}_{e,k}{\mathbf{\tilde{q}}} \!+\! 2\Re \{ {\mathbf{\tilde{q}}}^H\mathbf{\hat{H}}_{e,k}\Delta {\mathbf{q}}\} +{\mathbf{\tilde{w}}}^H\mathbf{\hat{H}}_{e,k}{\mathbf{\tilde{w}}} \\
\!\!\!\!\!  &+ 2\Re \{ {\mathbf{\tilde{w}}}^H\mathbf{\hat{H}}_{e,k}\Delta {\mathbf{w}}\}.
\end{split}
\end{equation}
According to \eqref{eq: zzy4} and \eqref{eq: zzy5}, we obtain linear approximations of
 the concave constraints \eqref{eq:Energy_constraint_user_PS} and \eqref{eq:Energy_constraint_eve} as
\begin{equation}\label{eq: zzy6}
\begin{split}
&{\mathbf{\tilde{q}}}^H\mathbf{H}_{c,l}{\mathbf{\tilde{q}}} \!+\! 2\Re \{ {\mathbf{\tilde{q}}}^H\mathbf{H}_{c,l}\Delta {\mathbf{q}}\} \!+\!
{\mathbf{\tilde{w}}}^H\mathbf{H}_{c,l}{\mathbf{\tilde{w}}}\\
&+ 2\Re \{ {\mathbf{\tilde{w}}}^H\mathbf{H}_{c,l}\Delta {\mathbf{w}}\} + \sigma_{c,l}^{2} \geq \frac{\bar{E}_{c,l}}{\eta_{c,l}(1-\rho_{c,l})}, ~\forall l,
\end{split}
\end{equation}
and
\begin{equation}\label{eq: zzy7}
\begin{split}
&{\mathbf{\tilde{q}}}^H\mathbf{\hat{H}}_{e,k}{\mathbf{\tilde{q}}} \!+\! 2\Re \{ {\mathbf{\tilde{q}}}^H\mathbf{\hat{H}}_{e,k}\Delta {\mathbf{q}}\} \!+\!{\mathbf{\tilde{w}}}^H\mathbf{\hat{H}}_{e,k}{\mathbf{\tilde{w}}} \\
&+ 2\Re \{ {\mathbf{\tilde{w}}}^H\mathbf{\hat{H}}_{e,k}\Delta {\mathbf{w}}\} \!+\! N_{R}\sigma_{k}^{2}  \geq \frac{\bar{E}_{e,k}}{\eta_{e,k}}, ~\forall k.
\end{split}
\end{equation}
Finally, by rearranging \eqref{eq:Power_constraints} as
 \begin{equation}\label{eq: zzy8}
\|[{\mathbf{q}}^T~~~{\mathbf{w}}^T]\|\leq \sqrt{P},
\end{equation}
the original problem \eqref{eq:Masked_beamforming_sec_rate_opt_ori} is transformed into
\begin{equation}\label{16}
\begin{split}
&\!\!\!\! \min_{\mathbf{q},{\kern 1pt}\mathbf{w},{\kern 1pt}\rho_{c,l}, {\kern 1pt}r_1, {\kern 1pt}r_2, {\kern 1pt}g_{r_1,l}, {\kern 1pt}g_{r_2,k}}  ~~~~~~ \|\mathbf{q}\| \\
&\!\!\!\! \mbox{s.t.} {\kern 3pt}\eqref{eq:11a}, {\kern 1pt}\eqref{eq:zzy8}, {\kern 1pt}\eqref{eq: zzy6}, {\kern 1pt}\eqref{eq: zzy7}, {\kern 1pt}\eqref{eq: zzy8}, 0 < \rho_{c,l} \leq 1, \forall l.
\end{split}
\end{equation}
Given $\mathbf{\tilde{q}}$, $\mathbf{\tilde{w}}$, $\tilde{r}_1$, and $\tilde{r}_2$, problem \eqref{16} is convex and can be efficiently solved by convex optimization software tools such as CVX \cite{28CVX}.
Based on the SPCA method, an approximation with the current optimal solution can be updated iteratively,
which implies that \eqref{eq:Masked_beamforming_sec_rate_opt_ori} is optimally solved.
In Section \ref{six}, we will show that the proposed SPCA method achieves the same performance as the 1-D search scheme, but with much lower complexity.

\section{Computational Complexity}
In this section, we evaluate the computational complexity of the proposed robust methods.
As will be shown in Section VI, the proposed SPCA algorithm achieves substantial improvement in complexity for the same performance compared with the method based on 1-D search.
Now we compare complexity of the algorithms through analyses similar to that in \cite{Complexity_zhu} \cite{TIFS_zhu}.
The complexity of the proposed algorithms are shown in Table I on the top of next page.
We denote $n$, $D = {{{\log }_2} {\frac{{{t _{\max }} - {t _{\min}}}}{\eta }} } $, and $Q$ as the number of decision variables, the 1-D search size, and the SPCA iteration number, respectively. The complexity analysis is given in the following.

\emph{1)}   PM with 1-D Search in problem \eqref{eq:relaxed_power_min_problem_results} involves $K$ LMI constraints of size $N_{{R}}+1$, two LMI constraints of size $N_{{T}}$,  and  $4L+K+1$ linear constraints. 

\emph{2)}  PM with SPCA in problem \eqref{16} has $L$ SOC constraints of dimension $2$, $L$ SOC constraints of dimension $N_{{T}}+1$, $K$ SOC constraints of dimension $2N_{{T}}+1$, one SOC constraints of dimension $2N_{{T}}$, and $L+3K$ linear constraints. 

For example, for a system with $L = 2, K = 3, N_T = 4,  N_R = 2$, $D = 100$, and $Q = 8$, the complexity of the PM with 1-D search, the PM with SPCA, are ${\cal O}(6.92\times 10^7)$, and  ${\cal O}(3.70\times 10^5)$, respectively. Thus, the complexity of the proposed SPCA method is only 1\% compared to the scheme based on 1-D search.

\begin{table}[htbp]
\caption{Complexity analysis of the proposed algorithms}
\label{tab:threesome}
\centering
\begin{tabular}{|c|c|}
\hline
 Algorithms & Complexity Order   \\
\hline
$\begin{array}{c}   {\textrm{1-D Search} } \\{{\textrm{method} }} \end{array}$ & $\begin{array}{l}  {{\cal O}\big( nD\sqrt {KN_R{\rm{+}}2K{\rm{+}}2N_T{\rm{+}}4L{\rm{+}}1}\big\{K(N_R{\rm{+}}1)^3} \\{{\rm{+}}2N_T^3{\rm{+}}n[K(N_R{\rm{+}}1)^2{\rm{+}}2N_T^2{\rm{+}}4L{\rm{+}}K{\rm{+}}1]{\rm{+}}n^2 \big\} \big)} \\{{\textrm{where}} ~ n ={\cal O}({2N^2_T} {\rm{+}} L)} \end{array} $ \\
\hline
$\begin{array}{c} {{\textrm{SPCA} }} \\{{\textrm{method} }} \end{array}$ & $\begin{array}{l}  {{\cal O}\big( nQ\sqrt {5K{\rm{+}}5L{\rm{+}}2}\big\{(2K{\rm{+}}L{\rm{+}}2)N_T{\rm{+}}3L{\rm{+}}K{\rm{+}}}\\{n(3K{\rm{+}}L){\rm{+}}n^2 \big\} \big)} ~{{\textrm{where}} ~ n ={\cal O}({2N_T} {\rm{+}} 2L{\rm{+}}K{\rm{+}}2)} \end{array} $ \\
\hline
\end{tabular}
\end{table}

\section{Numerical Results}\label{six}
In this section, we present numerical results to validate performance of the proposed transmit beamforming schemes.
 In the simulations, we consider a system  with $ N_{T} = 4 $,  $ L = 4 $, $ K = 3$, and $ N_{R} = 2$.
Both large-scale and small-scale fading are considered in the channel model. The simplified large-scale fading model is given by {$D_{L} = \big(\frac{d}{d_0}\big)^{-\alpha},$}
 where $d$ represents the distance between the transmitter and the receiver, $d_0$ is a reference distance equal to $10$ m in this work, and $\alpha= 3$ is the path loss exponent. We define $d_{c} = 40$ m as the distance between the transmitter and the CRs, and $d_e  = 20$ m as the distance between the transmitter and the ERs, unless otherwise specified.

{Because all the receivers are are expected to harvest energy from the RF signal, we consider line-of-sight (LOS) communication scenario where the Rician fading model is adopted for small scale fading coefficients.} 
The channel vector $\mathbf{h}_{c,l}$ is expressed as $\mathbf{h}_{c,l} = \sqrt{\frac{K_R}{1+K_R}}\mathbf{h}_{c,l}^{LOS} + \sqrt{\frac{1}{1+K_R}}\mathbf{h}_{c,l}^{NLOS}$,  where $\mathbf{h}_{c,l}^{LOS}$ indicates the LOS deterministic component with
$\| \mathbf{h}_{c,l}^{LOS} \|_2^2 = D_{L}$, $\mathbf{h}_{c,l}^{NLOS}$ represents the Rayleigh fading component as $\mathbf{h}_{c,l}^{NLOS} \sim \mathcal{CN}(0,D_{L}\mathbf{I})$, and $K_R=3$ is the Rician factor. It is noted that for the LOS component, we use the far-field uniform linear antenna array model \cite{30LOS_channel_model}.
In addition, we set $ \sigma_{c,l}^{2} = -60 ~\textrm{dBm}$, $ \sigma_{p,l}^{2} = -50 ~\textrm{dBm}$, $ \sigma_{k}^{2} = -50 ~\textrm{dBm}, \forall k$,
 $ \varepsilon_{c,l} = \varepsilon_{e,k} = \varepsilon$, $\mathbf{N}_{c,l} =\varepsilon^2 \mathbf{I}_{N_T}$, $\mathbf{D}_k = \varepsilon^2 \mathbf{I}_{N_TN_R}$, and $\eta_{c,l} = \eta_{e,k} =$ 0.3.

Fig. 1 illustrates the convergence of the SPCA method with respect to iteration numbers for $ P = 50$ dBm, $ \bar{E}_{c,l} = \bar{E}_{e,k} = E$, $R = 1$ bps/Hz, and $\varepsilon =$ 0.01. It is easily seen from the plots that convergence is achieved for all cases within just 8 iterations.

\begin{figure}[!htbp]
\centering
\includegraphics[scale = 0.5]{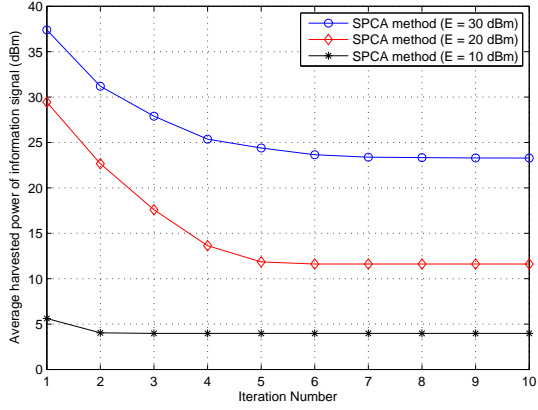}
\caption{Average transmit power of information signal versus iteration numbers}
\label{fig:power_VS_iternumber_161129}
\end{figure}

Fig. 2 illustrates the average transmit power of the information signal in terms of different target secrecy rates with $P = 30$ dBm and  $\bar{E}_{c,l}  = \bar{E}_{e,k} = 10$ dBm, $\forall k$. It is observed that the transmit power increases with the secrecy rate target. Here, the no-AN scheme is set with $\mathbf{W} = \mathbf{0}$.
In addition, the SPCA algorithm achieves the same performance as the 1-D search method, but with much lower complexity. Compared with the scheme without AN, the power consumption of the proposed AN-aided scheme is $9$ dB lower. Moreover, we can check that the proposed scheme performs better than the scheme with $\rho_{c} = \rho_{c,l} = 0.5$, and the performance gap becomes larger as the target secrecy rate increases. This indicates that optimizing the PS ratio $\rho_{c,l}$ is important, especially when the target secrecy rate is high.

\begin{figure}[!htbp]
\centering
\includegraphics[scale = 0.5]{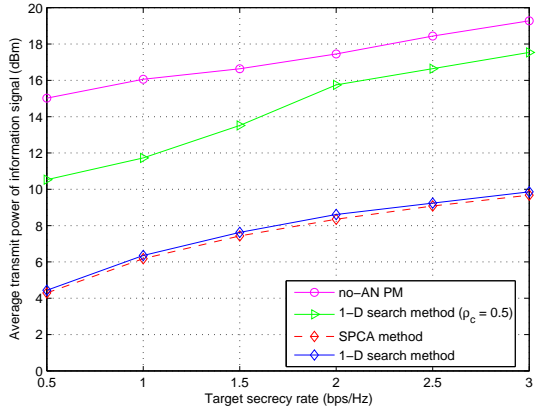}
\caption{Average transmit power of information signal versus target secrecy rate}
\label{fig:power_VS_rate_161129}
\end{figure}

\begin{figure}[!htbp]
\centering
\includegraphics[scale = 0.5]{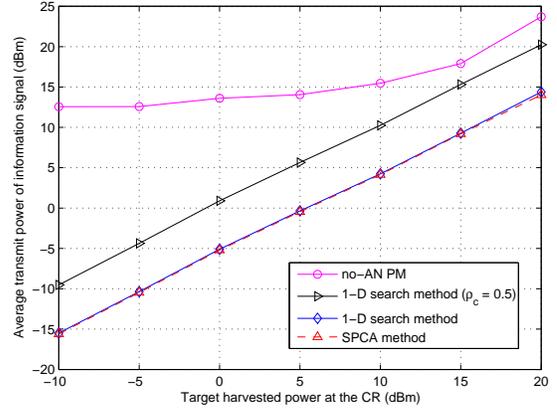}
\caption{Average transmit power of information signal versus the target harvested power at the CR}
\label{fig:power_VS_harvestedpower_161129}
\end{figure}

{In Fig. 3, we plot the average transmit power in terms of different target harvested power at the CR with $P = 40$ dBm, $\bar{E}_{e,k} = 10$ dBm and $\bar{R}_{c,l} = 0.5$ bps/Hz. We can check that the curves of the 1-D search method and SPCA method increase with the same slope.
Moveover, when the harvested power target decreases, the performance gap between the no-AN PM scheme and the proposed PM schemes become wider.
This indicates that AN is essential in achieving the performance gains. Furthermore, the 1-D search method require $6$ dB lower power than the 1-D search method with fixed $\rho_c$, respectively.}

\section{Conclusion}
In this paper, we have proposed AN-aided secure transmission scheme in multi-user MIMO SWIPT Systems where power splitters are employed by the receivers for SWIPT operation. The original problem, which was shown to be non-convex, was relaxed to formulate a two-layer problem.
The inner layer problem was recast as a sequence of SDPs. Then the optimal solution to the outer problem has been obtained through one-dimensional line search.
Moreover, tightness of the relaxation scheme has been investigated by showing that the optimal solution is rank-one.
To reduce the computational complexity, an SPCA based iterative algorithm has been proposed, which achieved near-optimal solution. Finally, numerical results have been provided to validate the performance of the proposed transmit beamforming schemes.

\appendices
\section{Proof of Theorem 1}
We first consider the Lagrange dual function of \eqref{eq:relaxed_power_min_problem_results} as
\begin{eqnarray}
&&\!\!\!\!\!\!\!\!\!\!\!\! \mathcal{L}(\mathbf{Q},\mathbf{W},\mathbf{Z},\mathbf{Y}, \xi_l, \mathbf{A}_{e,k},\gamma, \mu_l, \theta_{k}) \!=\! \textrm{tr}(\mathbf{Q})- \textrm{tr}(\mathbf{Z}\mathbf{Q})  - \textrm{tr}(\mathbf{Y}\mathbf{W}) \nonumber\\
&&\!\!\!\!\!\!\!\!\!\!\!\!- \xi_l\bigg[\textrm{tr}\bigg(\mathbf{h}_{c,l}\mathbf{h}_{c,l}^{H}[t \mathbf{Q} \!-\! (2^{\bar{R}_{c,l}} \!-\! t)\mathbf{W}]\bigg) - (2^{\bar{R}_{c,l}} - t)(\sigma_{c,l}^{2} \!+\! \frac{\sigma_{p,l}^{2}}{\rho_{c,l}})\bigg] \nonumber\\
&&\!\!\!\!\!\!\!\!\!\!\!\!- \sum_{k=1}^{K}\textrm{tr}\bigg[\mathbf{A}_{e,k}\bigg(\mathbf{H}_{e,k}^{H}\big(({\frac{1}{t}}  \!-\! 1)\mathbf{W} \nonumber -\mathbf{Q}\big)\mathbf{H}_{e,k} \!+\! ({\frac{1}{t}} \!-\! 1)\sigma_{k}^{2}\mathbf{I} \bigg)\bigg]  \nonumber  \\
&&\!\!\!\!\!\!\!\!\!\!\!\! + \gamma\big(\textrm{tr}(\mathbf{Q}\!+\!\mathbf{W}) \!-\! P\big)  \!-\! \mu_l\bigg[ \textrm{tr}\big(\mathbf{h}_{c,l}\mathbf{h}_{c,l}^{H}(\mathbf{Q}\!+\!\mathbf{W})\big)
\nonumber  - \frac{\bar{E}_{c,l}}{1\!-\!\rho_{c,l}} \!+\! \sigma_{c,l}^{2} \bigg] \nonumber  \\
&&\!\!\!\!\!\!\!\!\!\!\!\!- \sum_{k=1}^{K} \theta_{k}\bigg( \textrm{tr}\big(\mathbf{H}_{e,k}^{H}(\mathbf{Q}\!+\!\mathbf{W})\mathbf{H}_{e,k}\big)
- \bar{E}_{e,k} \!+\! N_{R}\sigma_{k}^{2} \bigg),\nonumber
\end{eqnarray}
where $ \mathbf{Z} \in \mathbb{H}_{+}^{N_{T}} $, $ \mathbf{Y} \in \mathbb{H}_{+}^{N_{T}} $, $ \xi_l \in \mathbb{R}_{+} $,
$ \mathbf{A}_{e,k} \in \mathbb{H}_{+}^{N_{T}} $, $ \gamma \in \mathbb{R}_{+} $,
$ \mu_l\in \mathbb{R}_{+} $, and $ \theta_{k} $ are the dual variables of $ \mathbf{Q} $, $ \mathbf{W} $,
\eqref{eq:objective_function_with_log_relaxed}, \eqref{eq:Eavesdroppers_rate_slack_variable_log_relaxed},
\eqref{eq:Power_constraints}, \eqref{eq:Energy_constraint_user_PS}, and \eqref{eq:Energy_constraint_eve}, respectively. Then, some of the related KKT conditions are listed as
\begin{subequations}
\begin{eqnarray}
&&\!\!\!\!\!\!\!\!\!\!\! \frac{\partial \mathcal{L}}{\partial \mathbf{Q}}   =
\mathbf{I} - \mathbf{Z} - (\xi_l t + \mu_l) \mathbf{h}_{c,l}\mathbf{h}_{c,l}^{H} +
\sum_{k=1}^{K}\mathbf{H}_{e,k}\mathbf{A}_{e,k}\mathbf{H}_{e,k}^{H} \nonumber \\
 &&~~~~ + \gamma \mathbf{I}  - \sum_{k=1}^{K}\theta_{k}\mathbf{H}_{e,k}\mathbf{H}_{e,k}^{H} = \mathbf{0}, \label{eq:Partial_derivation_Qs}\\
&&\!\!\!\!\!\!\!\!\!\!\! \frac{\partial \mathcal{L}}{\partial \mathbf{W}}   \!= \! -\mathbf{Y} \!+\!
[ \xi_l (2^{\bar{R}_{c,l}} \!-\! t) \!-\! \mu_l]\mathbf{h}_{c,l}\mathbf{h}_{c,l}^{H}  + \gamma\mathbf{I}  \!-\! \sum_{k=1}^{K}\theta_{k}\mathbf{H}_{e,k}\mathbf{H}_{e,k}^{H} \nonumber  \\
 &&~~~~ \!-\!
\sum_{k=1}^{K} ({\frac{1}{t}}\!-\!1)\mathbf{H}_{e,k}\mathbf{A}_{e,k}\mathbf{H}_{e,k}^{H}\!=\!
 \mathbf{0},\!\!\!\!\!\!\!\! \label{eq:Partial_derivation_W}\\
&&\!\!\!\!\!\!\!\!\!\!\! \mathbf{Z}\mathbf{Q}  = \mathbf{0}, ~\mathbf{Y} \!\succeq\! \mathbf{0},
~\mathbf{A}_{e,k} \succeq \mathbf{0}, ~ \xi_l \geq 0,~\mu_l\geq 0,~\forall k. \!\!\!\!\!\!\!\!\label{eq:Other_KKTs}
\end{eqnarray}
\end{subequations}

From the Lagrangian function and the KKT conditions, we have
 $ 0 < \rho_{c,l} \leq 1 $  and the KKT condition $ \xi_l > 0 $ and $ \mu_l> 0 $. Now, we will show these conditions via the dual
problem of \eqref{eq:relaxed_power_min_problem_results} as
\begin{eqnarray}\label{eq:Dual_problem_of_masked_beamforming_power_min}
&&  \!\!\!\!\!\!\!\!\!\! \max_{\mathbf{Z},\mathbf{Y},\mathbf{A}_{e,k},\xi_l,\gamma,\mu_l,\theta_{k}} \min_{\mathbf{Q},\mathbf{W},\rho_{c,l}}
 \mathcal{L}(\mathbf{Q},\mathbf{W},\mathbf{Z},\mathbf{Y}, \xi_l, \mathbf{A}_{e,k},\gamma,  \mu_l) \nonumber\\
=&&\!\!\!\!\!\!\!\!\!\!  \max_{\mathbf{Z},\mathbf{Y},\mathbf{A}_{e,k},\xi_l,\gamma,\mu_l,\theta_{k}} \min_{\mathbf{Q},\mathbf{W},\rho_{c,l}}
\bigg[- \textrm{tr}(\mathbf{Z}\mathbf{Q}) \!-\!  \textrm{tr}(\mathbf{Y}\mathbf{W}) \nonumber \\
&&\!\!\!\!\!\!\! +  \frac{\xi_l(2^{\bar{R}_{c,l}}\! -\! t)
\sigma_{p,l}^{2}}{\rho_{c,l}} \!+\! \frac{\mu_l\bar{E}_{c,l}}{1 \!-\!\rho_{c,l}} \!-\! \sum_{k=1}^{K}({\frac{1}{t}}\!-\! 1)\sigma_{e}^{2}\textrm{tr}
(\mathbf{A}_{e,k}) \\
&& \!\!\!\!\!\! + [\xi_l(2^{\bar{R}_{c,l}} \!-\! t) \!-\! \mu_l]\sigma_{c,l}^{2} \!-\! \gamma P  +
\sum_{k=1}^{K}\theta_{k}(\bar{E}_{e,k} \nonumber \! -  N_{R}\sigma_{k}^{2})\bigg].
\end{eqnarray}

Since problem \eqref{eq:relaxed_power_min_problem_results} is convex and satisfies the
Slater's condition, the duality gap between \eqref{eq:relaxed_power_min_problem_results} and
\eqref{eq:Dual_problem_of_masked_beamforming_power_min} is zero, and the strong duality
 holds. Therefore solving problem \eqref{eq:relaxed_power_min_problem_results} is
 equivalent to solving \eqref{eq:Dual_problem_of_masked_beamforming_power_min}.
In addition, the constraint $ 0 < \rho_{c,l} \leq 1 $ can be satisfied as
\begin{eqnarray}
\min_{0 < \rho_{c,l} \leq 1} \frac{\xi_l(2^{\bar{R}_{c,l}} \!-\! t)\sigma_{p,l}^{2}}{\rho_{c,l}} \!+\! \frac{\mu_l\bar{E}_{c,l}}{1\!-\!\rho_{c,l}}. \nonumber
\end{eqnarray}
Also the optimal variable $\rho_{c,l}^*$, and the dual variables
$\xi_l^*, \mu_l^*$ are related by
\begin{eqnarray}
 \rho_{c,l}^* \!=\! \frac{\sqrt{\xi_l^*(2^{\bar{R}_{c,l}} \!-\! t)\sigma_{p,l}^{2}}}{ \sqrt{\xi_l^*(2^{\bar{R}_{c,l}} \!-\! t)\sigma_{p,l}^{2}} \!+\! \sqrt{\mu_l^* \bar{E}_{c,l}}}.\nonumber
\end{eqnarray}

From the above inequality, we will show that $ \xi_l^* > 0 $ and $ \mu_l^* > 0 $ by
contradiction. Suppose that $ \xi_l^* = 0 $ and/or $ \mu_l^* = 0 $. Then there are two
cases (i.e., $ \rho_{c,l}^* = 0 $ or $ 1 $),
 which violate the constraints \eqref{eq:Achieved_sec_rate_constraint} and
 \eqref{eq:Energy_constraint_user_PS}. Thus, it follows $ \xi_l > 0 $ and $ \mu_l> 0 $. Now, subtracting \eqref{eq:Partial_derivation_W} from \eqref{eq:Partial_derivation_Qs} yields
\begin{eqnarray}\label{eq:Substraction_Z_Y}
 \mathbf{Z} \!+\! \xi_l^* 2^{\bar{R}_{c,l}} \mathbf{h}_{c,l}\mathbf{h}_{c,l}^{H} \!=\! \mathbf{I} \!+\! \mathbf{Y} \!+\! {\frac{1}{t}} \sum\nolimits_{k=1}^{K} \mathbf{H}_{e,k}\mathbf{A}_{e,k}\mathbf{H}_{e,k}^{H}.
\end{eqnarray}
We post-multiply $ \mathbf{Q} $ by both sides of \eqref{eq:Substraction_Z_Y} and use \eqref{eq:Other_KKTs} as
\begin{eqnarray}
\xi_l^* 2^{\bar{R}_{c,l}}\mathbf{h}_{c,l}\mathbf{h}_{c,l}^{H}\mathbf{Q} \!=\! \bigg(\! \mathbf{I} \!+\! \mathbf{Y} \!+\! {\frac{1}{t}} \sum\nolimits_{k=1}^{K} \mathbf{H}_{e,k}\mathbf{A}_{e,k}\mathbf{H}_{e,k}^{H} \!\bigg)\mathbf{Q}. \nonumber
\end{eqnarray}

Then, it becomes
\begin{eqnarray}
\xi_l^* 2^{\bar{R}_{c,l}}\big( \mathbf{I} \!+\! \mathbf{Y} \!+\! {\frac{1}{t}} \sum\nolimits_{k=1}^{K} \mathbf{H}_{e,k}\mathbf{A}_{e,k}\mathbf{H}_{e,k}^{H} \big)^{-1} \mathbf{h}_{c,l}\mathbf{h}_{c,l}^{H}\mathbf{Q} \!=\! \mathbf{Q}.\!\!\!\!\!\!\!\!\!\!\!\!\!\!\!\!\!\!\nonumber
\end{eqnarray}
Due to $ \xi_l^* > 0 $, we have
\begin{equation*}
\begin{split}
 &\textrm{rank}(\mathbf{Q})\nonumber \\
=& \textrm{rank}\bigg[ \xi_l^* 2^{\bar{R}_{c,l}} \bigg(\! \mathbf{I} \!+\! \mathbf{Y} \!+\! {\frac{1}{t}} \sum_{k=1}^{K} \mathbf{H}_{e,k}\mathbf{A}_{e,k}\mathbf{H}_{e,k}^{H} \!\bigg)^{-1} \mathbf{h}_{c,l}\mathbf{h}_{c,l}^{H}\mathbf{Q} \bigg]  \nonumber \\
=& \textrm{rank}(\mathbf{h}_{c,l}\mathbf{h}_{c,l}^{H}) \leq\! 1.
\end{split}
\end{equation*}

%


%
%
%

\end{document}